\newlength{\extralineskip}
\documentclass[12pt]{article}

\usepackage{amssymb}

\begin{document}
\begin{titlepage}
\begin{flushright}
          \begin{minipage}[t]{12em}
          \large UAB--FT--550\\
                 July 2003
          \end{minipage}
\end{flushright}
\vspace{\fill}

\vspace{\fill}

\begin{center}
\baselineskip=2.5em

{\large \bf Neutrino oscillations in the Sun probe long-range
leptonic forces}
\end{center}

\vspace{\fill}

\begin{center}
{\bf  J.A. Grifols and E. Mass\'o}\\
\vspace{0.4cm}
     {\em Grup de F\'\i sica Te\`orica and Institut de F\'\i sica
     d'Altes Energies\\
     Universitat Aut\`onoma de Barcelona\\
     08193 Bellaterra, Barcelona, Spain}
\end{center}
\vspace{\fill}

\begin{center}
\large Abstract
\end{center}
\begin{center}
\begin{minipage}[t]{36em}
Lepton number charges might be the source of long range forces. If
one accepts that neutrinos produced in the Sun do indeed oscillate
while crossing the interior of the Sun, then the shift in the
phase of the neutrino wavefunction caused by an hypothetical
potential associated to the leptonic charge of the electrons in
the Sun could affect the oscillation pattern beyond what is
actually observed. We show that a "fine structure" constant
$\alpha_{L}$ in excess of $6.4 \times 10^{-54}$ is incompatible
with present observational data. This bound is not valid for
forces whose range is shorter than the size of the Sun.
\end{minipage}
\end{center}

\vspace{\fill}

\end{titlepage}

\clearpage

\addtolength{\baselineskip}{\extralineskip}

A recurrent issue in
Particle Physics has been the question of long range fundamental
forces, notably, baryonic and leptonic forces. There is already a
considerable amount of literature \cite{dolgov} on these topics
that started with the seminal work by T.D. Lee and C.N. Yang on
long range forces coupled to baryonic charge \cite{lee}. From the
analysis of Eotvos type experiments, the hypothetical vector
bosons that mediate baryonic forces should couple to baryons with
a strength \cite{roll,brag}
\begin {equation}
\alpha_{B}<10^{-46}-10^{-47}
\end {equation}
where $\alpha_{B}$ is the corresponding "fine structure constant".
Similarly, Equivalence Principle tests that probe accelerations of
different elements towards the Sun, give a limit to the fine
structure constant associated to leptonic (electronic, indeed)
forces \cite{okun}
\begin {equation}
\alpha_{L}<10^{-48}-10^{-49}
\end {equation}
However, we now know that these electronic forces  cannot be
infinitely ranged because their associated vector bosons cannot be
strictly massless. A zero mass vector boson would imply exact
electronic number conservation and it is by now an established
fact that neutrinos oscillate \cite{bil} which implies violation
of lepton number. Still, if one insists that the associated
electronic forces extend over astronomical distances, the mass of
the vector boson must be very small. So, the constraint given
above (eq. (2)) should be valid for a "lepto-photon" with mass
less than or about $1.5 \times 10^{-18} $ eV (i.e. corresponding
to ranges larger or about 1 au.)

It is precisely the fact that neutrinos oscillate that we shall
exploit in the present paper to set an improved limit on the
electronic coupling constant in eq. (2). Laboratory limits on the
oscillation process have been used in the past to put constraints
on the Equivalence Principle or, conversely, putative violations
of the Principle of Equivalence have been suggested as a source of
the solar neutrino deficit \cite{gas,min,kras,mur,ray}. Now that
we know that the origin of the solar neutrino deficit is due to
mass related neutrino oscillations, we shall use this fact as a
handle to constrain an extra source of oscillation, namely that
due to long range leptonic forces. Indeed, if a force associated
to electron number does really exist then electron flavor
neutrinos transiting the interior of the Sun will feel the
influence of the electron number density in the Sun while muon
flavor neutrinos will not. This interaction, whose lepto-photons
should have a Compton wavelength on the order or larger than the
radius of the Sun, will induce a phase shift in the neutrino
propagation wave-function that should lead to neutrino
oscillations completely analogous to the way weak interactions
lead to oscillations in matter as first discussed by Wolfenstein
\cite{wolf,mik}.

Our starting standpoint is that neutrinos produced in the Sun
suffer large mixing angle (LMA) resonant MSW matter oscillations
with best-fit parameters given by \cite{fogl}:
\begin {equation}
\Delta m^2=5.5\times 10^{-5}{\rm eV}^2\, ,
\ \ \ \ \ \sin ^2 2\theta =0.83
\end {equation}
as a variety of different experimental inputs indicates. In the
case under consideration, the putative leptonic interaction adds a
piece in the hamiltonian that governs the time evolution of
neutrinos. In the flavor basis, this piece enters the interaction
hamiltonian as follows
\begin {equation}
\langle \nu_{e}\vert H_{int}\vert \nu_{e}\rangle =\sqrt2
G_{F}N_{e}+V_{L}
\end {equation}
and all other matrix elements vanish.

In eq. (4) $N_e$ is the electron number density and $V_{L}$ is the
potential energy of the neutrino in the field of the leptonic
force. It reads, for $\eta \lesssim R_{\odot}^{-1} $,
\begin {equation}
V_{L}(r)={\alpha_{L} \over r}\int _{0}^rd^3rN_{e}
\end {equation}
where $\eta$ is the lepto-photon mass and the corresponding Yukawa
potential has been approximated to a Coulomb-like one in the
distance range of interest. This interaction modifies the usual
neutrino-electron interaction length into
\begin {equation}
L_{e}=2\pi(\sqrt2 G_{F}N_{e}+V_{L})^{-1}
\end {equation}
The mixing angle in matter is given by the relation
\begin {equation}
\tan 2\theta_{m}=\tan 2\theta\, \left(1+{L_{V}\over L_{e}}\sec
2\theta \right)^{-1}
\end {equation}
where the vacuum oscillation length
\begin {equation}
L_{V}=2\pi (2E_{\nu}/\Delta m^2)
\end {equation}
In the conventional picture (i.e. for $\alpha_{L}\equiv 0$  ) and
for the values stated in eq. (3), resonant conversion occurs at
$20\%-30\%$ of the solar radius. Now, if we turn on gradually the
new interaction, the resonance region will move to thinner regions
of the Sun (i.e. further away from the solar center). This is due
to the repulsive character of the interaction that adds positively
to the weak potential.
Treating the potential (5) as a perturbation to the weak potential
we find, upon differentiation of the resonance condition
$L_{V}=L_{e}\cos 2\theta$,
\begin {equation}
{\delta \cos 2\theta \over \cos 2\theta}+{\delta \vert \Delta m^2
\vert \over \vert\Delta m^2 \vert}=2E_{\nu}V_{L}(r_{res})\sec
2\theta/\vert\Delta m^2 \vert
\end {equation}
i.e., for the perturbed potential the resonance condition is met
for slightly different oscillation parameters whose relative
variations are given by this formula. Now, we can feed the $95\%$
C.L. allowed deviations away from the oscillation parameters in
eq. (3) into the above relation to obtain the maximum $\alpha_{L}$
compatible with the data.
  To do the numerical work we
shall use a convenient parameterization of the electron number
density in the Sun, namely \cite{bac}
\begin {equation}
N_{e}/N_{A}=245\, e^{-10.54\, r/R_{\odot}}\, {\rm cm}^{-3}
\end {equation}
where $N_{A}$ is Avogadro's number. This fit to the number density
is not exact, particularly near the solar center (in the range
$0-0.17R_{\odot}$), but this fact is immaterial for our purposes
since i) the resonance position lies beyond $0.26R_{\odot}$ and
ii) $V_{L}$ itself vanishes in the central region.
So, using the above parameterization for the electron number
density and a mean neutrino energy of $10$ MeV (with these
inputs $r_{res}\simeq 0.27R_{\odot}$) we find from eq. (9) the
constraint
\begin {equation}
 \alpha_{L}\le 6.4 \times 10^{-54}.
\end {equation}
We have checked that the adiabaticity of the oscillations is
preserved when we incorporate the effects of the leptonic
potential with its maximum allowed strength given by eq. (11).
Indeed, we have scanned the whole region enclosed in the $95\%$
C.L. contour of the LMA domain and found that for no choice of the
parameters $\sin 2\theta$ and $\vert \Delta m^2 \vert$ the matter
oscillation length $L_{M}=L_{V}/\sin 2\theta$ at resonance exceeds
the width of the resonance in physical space. In fact, in all
instances explored, this latter quantity is much larger than
$L_{M}$. This guarantees that the neutrino can adjust itself to
the matter eigenstate while this latter slowly changes across the
resonance region. Furthermore, we verified explicitly that the
resonant transition is fully contained inside the Sun.

We remind the reader that the above limit on $\alpha_{L}$ is valid
only for $\eta \lesssim R_{\odot}^{-1}\sim 10^{-15} $ eV. It is
appropriate at this point to say that our phenomenological
approach to the issue of lepto-photon mass could run into serious
difficulties should this mass be too small \cite{okun1,okun2}.
Indeed, one might have to face catastrophic decay processes where
a huge number of longitudinal lepto-photons are radiated carrying
away the available energy (e.g., a muon decaying into an electron
and invisible energy). However, for the values of $\alpha_{L}$
obtained above (see eq. (11)) and $\eta \sim 10^{-15}$ eV, one can
easily verify using the results in ref. \cite{okun1,okun2} that we
are still very far from an infrared catastrophe. In fact, not even
one single longitudinal lepto-photon would be emitted in a muon to
electron transition.

To end this paper we should address the question whether screening
by the leptonic charges carried by electron neutrinos and
antineutrinos in the relict neutrino background affects the result
just derived. Indeed, the relict neutrinos in the cosmic plasma
might effectively screen the field created by the leptonic charge
associated to the electrons in the Sun and therefore invalidate
the bound given in equation (11) above. The problem of screening
of leptonic forces by cosmological neutrinos has been discussed in
the literature \cite {dolgov2,okun3}. The relevant quantity that
enters in those studies is the Debye length \cite{jac}, i.e. the
distance beyond which the screened potential effectively vanishes.
It is given by the expression
\begin {equation}
 l_{D}=\left (T_{\nu} \over 8\pi n_{\nu} \alpha_{L}\right )^{1/2}
\end {equation}
where $T_{\nu}\simeq 1.7 \times 10^{-4}$ eV and $n_{\nu}\simeq
115$ cm$^{-3} $ are the cosmic neutrino temperature and cosmic
electron neutrino density, respectively.   Even allowing for the
maximal strength of the coupling displayed in eq. (11) the Debye
screening length turns out to be on the order of ten Mpc, which is
an order of magnitude larger than the typical intergalactic
distance. Notice that  Debye screening can be physically relevant
only in the case where the range of the leptonic potential is
greater than $l_{D}$, otherwise the potential dies off before
screening takes over. Since, in our case, we need only the
potential to be felt by neutrinos over the size of the Sun, it is
clear that screening plays no role at solar radius ranges.

 One may wonder whether $l_{D}$  could we
affected by the fact that electronic charge is not conserved, i.e.
whether a cosmological lepton asymmetry obtains as a result of
neutrino oscillations. This is however not the case because ever
since neutrino decoupling, i.e. when neutrinos had their last weak
interaction, electron neutrinos have been oscillating into muon
neutrinos and muon neutrinos into electron neutrinos at the same
pace. Hence, in a comoving volume the net leptonic (electronic)
charge is conserved because as much electron number is destroyed
as it is created.

{\it Note added:}
After this paper was completed we learned that related work
using atmospheric neutrinos has been performed in 
\cite{Joshipura:2003jh}. Work linking, from a completely
different perspective, long-range forces with the
MSW effect in the Sun can be found in \cite{Horvat:mt}.

Work partially supported by the CICYT Research Project
FPA2002-00648, by the EU network on Supersymmetry and the Early
Universe HPRN-CT-2000-00152, and by the DURSI Research Project
2001SGR00188.

\end{document}